\documentclass[preprint]{revtex4-1}
\usepackage{amsmath}
\usepackage{amsfonts}
\usepackage{graphicx}
\usepackage{caption}

\begin{document}

\title{Single particle energy diffusion from relativistic spontaneous localization}
\date{\today}
\author{D. J. Bedingham}
\affiliation{Blackett Laboratory \\ Imperial College \\ London SW7 2BZ \\ UK}

\begin{abstract}
Energy diffusion due to spontaneous localization (SL) for a relativistically-fast moving particle is examined. SL is an alternative to standard quantum theory in which quantum state reduction is treated as a random physical process which is incorporated into the Schr\"odinger equation in an observer-independent way. These models make predictions in conflict with standard quantum theory one of which is non conservation of energy. On the basis of proposed relativistic extensions of SL it is argued that for a single localized particle, non-relativistic SL should remain valid in the rest frame of the particle. The implication is that relativistic calculations can be performed by transforming non-relativistic results from the particle rest frame to the frame of an inertial observer. This is demonstrated by considering a relativistic stream of non-interacting particles of cosmological origin and showing how their energy distribution evolves as a result of SL as they traverse the Universe. A solution is presented and the potential for astrophysical observations is discussed.
\end{abstract}

\maketitle

\section{Introduction}
Motivated by the measurement problem, spontaneous localization (SL) models are an alternative to standard quantum theory in which quantum state reduction is treated as a genuine physical process \cite{REP1,REP2}. The typical formulation of these models is by modification of the Schr\"odinger equation to include non-linear and stochastic terms. Both of these features are well motivated. Stochasticity represents the random nature of state reduction. Non linearity enters since the probability of the state reducing to a particular outcome depends on the state itself (i.e.~the Born rule). This approach is an empirical  way of modelling the behaviour of a quantum state as it is observed to behave in practice whether in a measurement situation or not.  

The state-of-the-art non-relativistic SL model is the continuous spontaneous localization (CSL) model \cite{CSL1,CSL2}. This model is formulated in terms of a stochastic differential equation for the state vector (see below) describing a continuous stochastic state trajectory in Hilbert space. The CSL model reproduces the behaviour of non-relativistic quantum systems on the micro scale whilst any macro superpositon of quasi-localized states is rapidly suppressed (the position basis takes a special role). This happens without having to make an arbitrary division between the micro and the macro domains - the theory itself determines when a superpostion state is stable and when it is not. The remarkable thing is that this works in a way which is consistent with our experience.

This paper concerns a side effect of SL models which is that they lead to a gradual increase in the energy of a system on average. There are two complimentary ways in which this happens. The first is that as a result of the localization process, an initially spread out wave function becomes narrower in position space and therefore broader in momentum space. Since the free Hamiltonian is a convex function of momentum, the expected energy increases. The second contribution is due to the fact that as the localizations happen the wave packet as a whole tends to undergo stochastic jumps in phase space (this is examined in more detail below). This also results in spreading of momentum and consequently increases the expected energy on average.

We shall consider this effect when the particle is travelling at relativistic speed with respect to some observer. For the sake of definiteness we consider a stream of relativistically-fast non-interacting particles whose origin might be in the early universe. These particles travel at close to the speed of light and spend almost the whole lifetime of the Universe travelling freely until they eventually collide with an observer on Earth. The question we wish to ask is this: Would the non conservation of energy due to SL lead to a measurable effect in the energy distribution of the observed particles?

An obstacle in getting to this goal is that existing relativistic collapse models are complicated. In particular both the relativistic models of Refs\cite{rel1} and \cite{rel2} are non Markovian and this makes calculations more difficult to perform. We shall therefore take a short cut, arguing that the non-relativistic SL equations for a single particle should be valid in its rest frame. 

A key feature of SL is that the localizations have an associated length scale. This is clear in discrete models such as the GRW model \cite{GRW} where the state occasionally and randomly collapses under the action of a Gaussian quasi projector in position space. The Gaussian has an associated length scale defining its spread in space. This is essential. Were we to try to remove the length scale by making the localizations infinitely sharp the resulting collapsed wave function would have infinite energy. The same is true in CSL. 

To retain the feature of a fundamental length scale without resorting to the use of a preferred frame or foliation has proved to be one of the main difficulties when formulating relativistic extensions of SL. For example, in Ref.\cite{rel1} the localizations are centred about a random, time ordered sequence of points in spacetime, called flashes. The state at any given stage is defined on an arbitrary spacelike surface to the future of the previous flash. The law of the flashes requires that the state is unitarily evolved to a hyperbolic surface a random (Poisson distributed) timelike distance to the future of the previous flash; the state then defines a probability distribution on the hyperbolic surface from which the location of a new flash is randomly drawn. The state is then modified by the action of a Gaussian quasi projector with fixed length scale acting within the hyperbolic surface and centred on the flash location. We can deform the spacelike surface to the future of this new flash - the state evolves unitarily until the next random flash happens. This procedure is independent of any particular frame of reference. 

Now consider two observers - one in the rest frame of the particle (which we take to be well localized) and another moving with respect to it. For the rest frame observer, provided that the proper time between flashes is sufficiently large compared to the length scale of the localization operator, the hyperboloid can be approximated by a plane and the model reduces to the non-relativistic GRW model. However, due to the invariance of the construction, the moving observer will simply see the localization length contracted and the time between flashes dilated. 

A further example is provided by the model of Ref.\cite{rel2}. This is a model involving quantum fields in which a new spacetime quantum field is introduced to mediate the influence of the localizations. The length scale arises from a smeared interaction between quantum fields and the mediating field. In order to avoid divergences, the smearing must be confined to a finite spacetime region. This is achieved by allowing the smearing function to depend on local properties of the quantum fields. The construction proposed results in a smearing function which defines a region in spacetime that is near to the point of interaction from the point of view of an averaged local rest frame.

In each of these relativistic models the localizations can be said to happen with a fixed length scale in the rest frame of the system (at least in the case of a well localized particle which we consider here where the rest frame is unambiguous). It seems natural on general grounds that this should be a universal feature of relativistic collapse models. A length scale is necessary in order to define a localization, therefore a frame in which the length scale applies is necessary. The obvious way to do this without making reference to a preferred foliation is to make use of local rest frames defined by the state. Whilst it might not be obvious precisely how to do this for a general state (perhaps requiring a fairly technical definition as in both the relativistic models mentioned above), for a single particle in a localized state the rest frame is clear. 

Relativistic invariance ensures that the localization process seen from a moving observer's point of view is simply a Lorentz transformation of events. We therefore take the approach of working in the particle rest frame where non-relativistic equations of motion are adequate, before transforming to the moving observers frame. Note that this is an assumption inspired by relativistic considerations. An alternative possibility which we will not explore here is that there exists a preferred global frame (e.g.~the cosmological frame) in which the localizations occur. In this case we might make the approximation that the non-relativistic CSL equations hold in only in this special frame. 

The structure of the paper is as follows: In Section \ref{S2} we show how the CSL model can be significantly simplified in the case of a state describing a single localized particle. We describe the steady state solution for the CSL model in this limit and reduce the behaviour of the system to a classical diffusion in phase space. In Section \ref{S3} we derive the energy diffusion process and perform a transformation from the rest frame of the particle to the frame of an inertial observer moving at relativistic speed with respect to the particle. We solve this diffusion equation to find the probability distribution for the particle to end up with a given energy. We conclude with some discussion in Section \ref{S4}.

\section{CSL in the localized single particle limit}
\label{S2}

In this section we organize the necessary non-relativistic results taken to hold in the rest frame of the particle. We will demonstrate the steady state solution for the CSL model in the case of a single, well localized particle. In order to do this we first demonstrate that the CSL equations can be simplified to a form known as QMUPL (quantum mechanics with universal position localization) \cite{DIO2}. It is well known that relationships such as this exist between the various SL models in certain limits \cite{DIO2,SIM2, SIM3, CSL2}. However, the following demonstration is believed to be novel.

The CSL model concerns a quantum system described in terms of a non-relativistic quantum field. The state evolution is described by the stochastic differential equation
\begin{align}
d|\psi\rangle =& \left[-\frac{i}{\hbar} \hat{H}- \frac{\lambda}{2}\int d{\bf x} \left(\hat{N}({\bf x}) - \langle \hat{N}({\bf x}) \rangle\right)^2\right]dt |\psi\rangle
\nonumber\\
&+ \sqrt{\lambda}\int d{\bf x}\left(\hat{N}({\bf x}) - \langle \hat{N}({\bf x}) \rangle\right) dB_t ({\bf x})|\psi\rangle,
\label{CSL}
\end{align}
where the number density operator $\hat{N}({\bf x})$ is given by
\begin{align}
\hat{N}({\bf x}) = \left( \frac{\alpha}{\pi}\right)^{3/4}\int d{\bf x}\exp\left\{-\frac{\alpha }{2}({\bf x}-{\bf y})\cdot({\bf x}-{\bf y})\right\}\hat{a}^{\dagger}({\bf y})\hat{a}({\bf y}),
\end{align}
the field annihilation and creation operators $\hat{a}({\bf x})$ and $\hat{a}^{\dagger}({\bf x})$ satisfy
\begin{align}
[\hat{a}({\bf x}),\hat{a}^{\dagger}({\bf y})] = \delta({\bf x}-{\bf y}),
\end{align}
and the field of Brownian motions satisfy
\begin{align}
\mathbb{E}[dB_t({\bf x})] = 0;\quad dB_s({\bf x})dB_{t}({\bf y}) = \delta_{st}\delta({\bf x}-{\bf y})dt.
\end{align}
The CSL parameters $\lambda$ and $1/\sqrt{\alpha}$ are understood respectively as the rate and the length scale of localization. The state remains normalized under Eq.(\ref{CSL}) and we note that the Schr\"odinger equation is recovered in the limit that $\lambda\rightarrow 0$. As mentioned in the Introduction this model gives close agreement with the standard Schr\"odinger equation for micro systems but leads to rapid suppression of macro superpositions of quasi-localized states.

Our task now is to simplify this construction for the purposes of our calculation. We first assume that the state is describing a single particle. The approximation will be valid also if we a considering a swarm of non-interacting, non-overlapping, and non-entangled particles where the state factorizes into single particle states. For a single particle the state is represented by
\begin{align}
|\psi\rangle = \int d{\bf x}\; \psi({\bf x}) \hat{a}^{\dagger}({\bf x})|0\rangle,
\end{align}
where $|0\rangle$ is the vacuum state. Here we can identify $\psi({\bf x})$ as the wave function for the particle. Improper position eigenstates take the form $|{\bf x}\rangle = \hat{a}^{\dagger}({\bf x})|0\rangle$ and the position operator is given in terms of field creation and annihilation operators by
\begin{align}
\hat{\bf x} = \int d{\bf x}\; {\bf x} \hat{a}^{\dagger}({\bf x})\hat{a}({\bf x}).
\end{align}
Given this definition we find
\begin{align}
\langle \hat{\bf x}\rangle = \int d{\bf x}\; {\bf x} |\psi({\bf x})|^2,
\end{align}
as expected.

Next we assume that the particle is sufficiently localized about a point $\bar{{\bf y}}$ that we can make the approximation (cf.~Appendix B of \cite{SIM1})
\begin{align}
\exp\left\{-\frac{\alpha}{2}({\bf x}-{\bf y})\cdot ({\bf x}-{\bf y})\right\}\simeq 
\exp\left\{-\frac{\alpha}{2} ({\bf x}-\bar{{\bf y}})\cdot ({\bf x}-\bar{{\bf y}})\right\}\left[1+\alpha ({\bf x}-\bar{{\bf y}})\cdot ({\bf y} -\bar{{\bf y}})\right]
\end{align}
This requires that $|{\bf y}-\bar{{\bf y}}|\ll 1/\sqrt{\alpha}$, i.e.~the particle must be localized about the point $\bar{\bf y}$ on a length scale much smaller than the localization length scale of the CSL model. The point $\bar{{\bf y}}$ is time dependent - it describes the location of the centre of the particle's wave packet.

We can combine these various elements to calculate the following useful relations
\begin{align}
&\int d{\bf x} \hat{N}^2({\bf x}) \simeq 1 + \frac{\alpha}{2}\bar{\bf y}\cdot \bar{\bf y} +\frac{\alpha}{2}\hat{\bf x}\cdot \hat{\bf x} -\alpha\bar{\bf y}\cdot \hat{\bf x}, \nonumber\\
&\int d{\bf x} \langle \hat{N}({\bf x})\rangle^2 \simeq 1 + \frac{\alpha}{2}\bar{\bf y}\cdot \bar{\bf y} +\frac{\alpha}{2}\langle \hat{\bf x}\rangle \cdot \langle \hat{\bf x} \rangle-\alpha\bar{\bf y}\cdot \langle \hat{\bf x} \rangle,\nonumber\\
&\int d{\bf x} \langle \hat{N}({\bf x})\rangle \hat{N}({\bf x}) \simeq 1 + \frac{\alpha}{2}\bar{\bf y}\cdot \bar{\bf y} +\frac{\alpha}{2}\langle \hat{\bf x}\rangle \cdot \hat{\bf x} 
-\frac{\alpha}{2}\bar{\bf y}\cdot \langle \hat{\bf x} \rangle-\frac{\alpha}{2}\bar{\bf y}\cdot \hat{\bf x}, \nonumber\\
& \int d{\bf x} \hat{N}({\bf x}) dB_t({\bf x}) \simeq d\tilde{B}_t + \sqrt{\frac{\alpha}{2}}(\hat{\bf x}-\bar{\bf y}) \cdot d{\bf B}_t, \nonumber\\
& \int d{\bf x} \langle \hat{N}({\bf x}) \rangle dB_t({\bf x}) \simeq d\tilde{B}_t +  \sqrt{\frac{\alpha}{2}}(\langle \hat{\bf x} \rangle-\bar{{\bf y}}) \cdot d{\bf B}_t,
\label{relations} 
\end{align}
where the Brownian motions ${\bf B}_t$ and $\tilde{B}_t$ are related to the Brownian motion field through
\begin{align}
& d{\bf B}_t = \int d{\bf x} \sqrt{2\alpha} \left(\frac{\alpha}{\pi}\right)^{3/4}\exp\left\{-\frac{\alpha}{2}({\bf x}-\bar{\bf y})\cdot ({\bf x}-\bar{\bf y})\right\} ({\bf x}-\bar{\bf y})dB_t({\bf x}),\\
& d\tilde{B}_t = \int d{\bf x} \left(\frac{\alpha}{\pi}\right)^{3/4}\exp\left\{-\frac{\alpha}{2}({\bf x}-\bar{\bf y})\cdot ({\bf x}-\bar{\bf y})\right\} dB_t({\bf x}).
\end{align}
These definitions are readily shown to satisfy
\begin{align}
\mathbb{E}[dB_{i,t}] = 0 = \mathbb{E}[d\tilde{B}_t];\quad dB_{i,s} dB_{j,t} = \delta_{ij}\delta_{st}dt; \quad d\tilde{B}_s d\tilde{B}_{t} = \delta_{st}dt; \quad dB_{i,s} d\tilde{B}_{t} = 0,
\label{browns}
\end{align}
where $dB_{i,t}$ for $i = 1,2,3$ are the three orthogonal components of $d{\bf B}_t$ relating to the orthogonal spatial directions $x_i$.

Inserting the relations (\ref{relations}) into Eq.(\ref{CSL}) results in 
\begin{align}
	d|\psi\rangle = \left\{-\frac{i}{\hbar}\hat{H} dt  - \frac{D}{\hbar^2}( \hat{\bf x}-\langle \hat{\bf x}\rangle)\cdot ( \hat{\bf x}-\langle \hat{\bf x}\rangle) dt 
+ \frac{\sqrt{2D}}{\hbar}(\hat{\bf x}-\langle \hat{\bf x} \rangle)\cdot  d{\bf B}_{t}\right\}|\psi\rangle,
\label{QSD}
\end{align}
with $D$ given in terms of the CSL parameters as
\begin{align}
D =\frac{\lambda \alpha \hbar^2}{4}, 
\label{D0}
\end{align}
(the factor of $\hbar^2$ is included for later convenience). In terms of momentum space field creation and annihilation operators the Hamiltonian is given by 
\begin{align}
\hat{H} = \int d{\bf p} \; \frac{{\bf p}\cdot {\bf p}}{2m}\hat{a}^{\dagger}({\bf p}) \hat{a}({\bf p}),
\end{align}
so that $\langle {\bf p}|\hat{H}|\psi\rangle = ({\bf p}\cdot {\bf p}/2m)\psi({\bf p})$, where $|{\bf p}\rangle =\hat{a}^{\dagger}({\bf p})|0\rangle$, and $\psi({\bf p}) = \int dx e^{-i{\bf p}\cdot {\bf x}/\hbar}\psi({\bf x})/\sqrt{2\pi\hbar}$. We can therefore write
\begin{align}
\hat{H} = \frac{\hat{\bf p}\cdot\hat{\bf p}}{2m},
\end{align}
in the single particle case.

Equation (\ref{QSD}) is the QMULP model \cite{DIO2}. Not only is this equation useful as a scaled limit of CSL, it can also be used to describe the effect of a thermal environment when dealing with open systems (see e.g.~\cite{ENV}).

The advantage of Eq.(\ref{QSD}) for our purpose is that we can calculate the steady state limit which occurs when the diffusive effects of the Hamiltonian balance with the localizing effects of the SL terms \cite{DIO1,SIM2}. The form of the wave function in the steady state is given by
\begin{align}
\psi({\bf x}) = \frac{1}{(2\pi\sigma_{\infty}^2)^{3/4}}
\exp\left\{-\frac{(1-i)}{4\sigma_{\infty}^2}({\bf x} - \langle \hat{\bf x} \rangle)\cdot ({\bf x} - \langle \hat{\bf x} \rangle)+\frac{i}{\hbar}\langle \hat{\bf p}  \rangle \cdot{\bf x}\right\}, 
\end{align}
where the steady state width is
\begin{align}
\sigma_{\infty} = \sqrt[4]{\frac{\hbar^3}{8 D m}}.
\end{align}
Starting from a general wave function it takes a time of order $t_{\rm loc}\sim\sqrt{m\hbar/D}$ to reach the steady state. The shape of the wave function is stable, however, the average position and momentum of the packet undergoes 3D Brownian motion described by the following diffusion equations
\begin{align}
	d\langle \hat{\bf x} \rangle &= \frac{\langle \hat{\bf p} \rangle}{m}dt + \sqrt{\frac{\hbar}{m}} d{\bf B}_{t} 
	\label{x}\\
	d\langle \hat{\bf p} \rangle &=  \sqrt{2D} d{\bf B}_{t} .
	\label{p}
\end{align}
These closed equations describe a classical Brownian motion for the wave packet.

We now exhibit estimates of the steady state widths and localization times for a set of different species of particles. A detailed analysis of the valid range of choices for the SL parameters consistent with experiments can be found in Ref.\cite{TUM2}. Here we use the original values suggested by GRW \cite{GRW} of $\lambda \sim 10^{-16}s^{-1}$ and $1/\sqrt{\alpha} \sim 10^{-7}m$. Based on experimental evidence of spontaneous photon emission rates from Germanium it has been possible to argue that $\lambda$ should increase with the mass of the particle as $m^2$ \cite{M2}. This result we put in by hand (we would expect it to ultimately result from a more fundamental theory, perhaps involving gravity). We therefore assume that the GRW value relates to a single nucleon (as originally intended) and write
\begin{align}
\lambda = \left(\frac{m}{m_n}\right)^2\times 10^{-16}s^{-1},
\label{lamb}
\end{align}
 where $m_n$ is a nucleon mass. Based on these assumptions the results are shown in the following table:
\begin{center}
	\begin{tabular}{|l|c|c|}\hline
		Particle & $\sigma_{\infty}$ & $t_{loc}$ \\ \hline
	    	neutrino ($0.1eV/c^2$) & $1300km$ & $180yrs$\\ 
		electron &  $12m$ & $41days$\\  
		proton & $4cm$ & $23hrs$\\  
		Fe nucleus & $2mm$ & $3 hrs$\\	
	    	10,000u cluster & $42\mu m$ &  $14 mins$\\  \hline
	\end{tabular}
\captionof{table}{Steady state widths and localization times for various types of particle based on the GRW parameters.}
\label{T1}
\end{center}
We see from Table \ref{T1} that none of the particles satisfy the condition that $\sigma_{\infty}\ll 1/\sqrt{\alpha}$ for the GRW parameters. However, we note that many results relating to the effects of spontaneous localization involve the parameters $\lambda$ and $\alpha$ only in the combination $\lambda\alpha$ (e.g. reduction rates, rate of energy increase). We therefore assume that the combination $\lambda\alpha$ takes the GRW value (for a single nucleon) and that $\alpha$ is as large as necessary to fulfil the localized particle approximation made in this section.

For a proton the value for $\sigma_{\infty}=4cm$ gives a sense of being fairly well localized on large scales such that it can be treated like a particle, whilst at the same time very spread out on atomic scales such that its wavelike characteristics should dominate.

\section{Relativistic CSL}
\label{S3}
The localization times for the various particles in Table \ref{T1} indicate that on cosmological time scales a free particle can be expected to exist in its steady state. We therefore assume that the particle is in its steady state and concern ourselves only with the classical diffusive motion of the packet. Since the behaviour is reduced to a classical form we write $\langle \hat{\bf x}\rangle = {\bf X}$ and  $\langle \hat{\bf p}\rangle = {\bf P}$. We assume that Eqs(\ref{x}) and (\ref{p}) apply in the rest frame of the particle - defined by ${\bf P} = 0$ - whereby relativistic effects can be ignored. Note that due to the diffusive motion of the particle the rest frame is continuously changing. Below we describe how to transform this process to the frame of an inertial observer. 

Denoting the rest frame ${\cal O}'$ and labelling coordinates in this frame with a prime we have from Eqs(\ref{x}) and (\ref{p})
\begin{align}
	d{\bf X} ' &= \sqrt{\frac{\hbar}{m}} d{\bf B}_{t'} 
	\label{x2}\\
	d{\bf P}' &=  \sqrt{2D} d{\bf B}_{t'} .
	\label{p2}
\end{align}
We assume that the energy of the particle near its rest frame is given by the non-relativistic formula $E' = {\bf P}' \cdot {\bf P}'/2m$. Using It\^o's lemma and (\ref{browns}) we can derive the process for the energy
\begin{align}
dE' = \frac{3D}{m}dt' +\frac{\sqrt{2D}}{m}{\bf P}'\cdot  d{\bf B}_{t'}.
\label{dE}
\end{align}
In the rest frame of the particle this is simply given by
\begin{align}
dE' = \frac{3D}{m}dt'.
\end{align}
The drift in energy is due to the stochastic shifts in momentum described by Eq.(\ref{p2}).

We now transform from the particle rest frame to the cosmological frame ${\cal O}$ (coordinates in this frame are unprimed) assumed to be an inertial frame in which the particle travels at relativistic speed $v\sim c$ (we refer to this as the ultra-relativistic limit) in the $X_i$ direction. This direction will change as the particle undergoes diffusion - we rotate coordinates accordingly.  The energy process in the cosmological frame is given by the Lorentz transformation
\begin{align}
dE &= \gamma dE' +v \gamma d {P}_{i} ' \nonumber \\
&\simeq \gamma dE' +c \gamma d{P}_{i}' \nonumber \\
& = \frac{3D}{m}\gamma dt' +\sqrt{2D} c\gamma dB_{i,t'},
\label{dE1}
\end{align}
with $\gamma = 1/\sqrt{1-v^2/c^2}$. This process is still expressed in terms of the rest frame time $t'$. We would like to express it in terms of the cosmological time $t$. To do this we use the Lorentz transformation
\begin{align}
	dt = \gamma\left(dt' - \frac{vd {X}_i'}{c^2}\right) \simeq \gamma\left(dt' - \frac{d {X}_i'}{c}\right).
	\label{time}
\end{align}
Depending on the random fluctuations which define $d{X}_i'$, $dt$ can be positive or negative. This indicates that the particle in fact moves in tiny spacelike jumps. This makes the definition (\ref{time}) unsuitable as the time parameter. A suitable definition can be guessed by considering a finite period of cosmological time
\begin{align}
	t = \int_0^{t'}\gamma ds' - \sqrt{\frac{\hbar}{mc^2}}\int_0^{t'}\gamma d B_{i,s'}.
\end{align}
Note that $\gamma$ is a function of $t'$ (since the velocity $v$ is a function of $t'$). The two terms on the right hand side can be thought of as a signal term and a noise term (with expectation zero). Then for a given value of $t'$ we can estimate $t$ by
\begin{align}
	t= \int_0^{t'}\gamma ds'.
\end{align}
The standard deviation in this estimate is given by the standard deviation of the It\^o integral term. This is given by
\begin{align}
	\sigma_t 	= \left(\frac{\hbar}{mc^2}\int_0^{t'}\mathbb{E}[\gamma^2]ds'\right)^{1/2}.
\end{align}
So for $\gamma$ approximately constant corresponding to a small amount of velocity dispersion we can write $\mathbb{E}[\gamma^2]\sim\gamma^2$ and the condition that $\sigma_t \ll t $ corresponds to
\begin{align}
	\frac{\hbar}{mc^2}\ll t'.
\end{align}
For a proton $\hbar/mc^2\sim 10^{-24}s$ so on cosmological time scales we can safely ignore the stochastic shifts of position of the wave packet in its rest frame.

In order to convert to cosmological time we therefore write $dt = \gamma dt'$ and using a theorem of stochastic calculus we have $dB_{t} = \gamma^{1/2}dB_{i,t'}$ with $dB_{t}^2 = dt$ (we drop the $i$ since there is only one Brownian motion factor in the energy process). Using $E = \gamma mc^2$, Eq.(\ref{dE1}) can now be written
\begin{align}
	dE = \frac{3D}{m} dt +\sqrt{\frac{2D}{m}E} \;dB_{t}.
	\label{E}
\end{align}
It is worth noting that this equation is remarkably similar to the non-relativistic equation that is obtained from equation (\ref{dE}). We find that the only difference is an extra factor of $\sqrt{2}$ in the stochastic term of the non-relativistic version.

For generality we include the effects of the expansion of the Universe as a friction term resulting in
\begin{align}
	dE = \left\{\frac{3D}{m} -\frac{\dot{a}}{a}E\right\}dt+\sqrt{\frac{2D}{m}E} \; dB_{t},
	\label{dE3}
\end{align}
where $a$ is the scale factor of the Universe and $\dot{a}/a$, the Hubble parameter, is taken to be a constant. This equation describes the diffusion in energy of a particle due to CSL as it travels at relativistic speed with respect to a cosmological observer (subject to the various approximations outlined earlier).

Eq.(\ref{dE3}) is in fact a Cox-Ingersoll-Ross (CIR) process \cite{CIR}. The process is well known in finance where it is used to describe an instantaneous interest rate. A generic process of this type
\begin{align}
dz = \kappa(\theta - z)dt +\sigma\sqrt{z}dB_t,
\end{align}
has the properties that $z$ is elastically pulled towards the long-term value $\theta$ at a rate determined by $\kappa$; the origin is inaccessible if $2\kappa\theta \geq \sigma^2$ in which case $z$ remains positive; the variance in $z$ increases as $z$ increases; and there is a steady state distribution for $z$.

How do these properties apply to the problem of relativistic energy diffusion? The condition for the origin to be inaccessible is satisfied meaning that the energy will always remain positive. We might expect that the energy would have a long term average value given by $(3D/m)/(\dot{a}/a)$ (independent of the initial state). However, the rate at which $E$ is elastically pulled to this value is $\dot{a}/a$ - of order of the inverse age of the Universe. Therefore for the energy to reach its steady state distribution will require a time much longer than the age of the Universe (or more correctly, a much longer time than the time over which the Hubble parameter can be approximated as a constant). 

The forward equation corresponding to Eg.(\ref{dE3}) is given by 
\begin{align}
\frac{d}{dt}p_t(E|E_0) = \left\{ \frac{D}{m} E^2\frac{\partial^2}{\partial E^2}
				+\left(\frac{\dot{a}}{a}  - \frac{D}{m}\right)\frac{\partial}{\partial E}
				+ \frac{\dot{a}}{a}\right\}p_t(E|E_0).
\label{FWD}
\end{align}
We can compare this equation with a result from Refs.\cite{SW1,SW2} where a diffusion process in phase space due to a fundamental discreteness of spacetime is considered. There it is shown that starting from the idea of a random walk in momentum space, the condition of Lorentz invariance alone can be used to determine the form of the equation satisfied by the probability distribution on phase space. It turns out that Eq.(\ref{FWD}) is consistent with this result. In order to show this we find the equation satisfied by the marginal distribution for momentum and then convert from momentum to energy in the ultra-relativistic limit. This gives us confidence that our result (\ref{FWD}) is at least relativistically correct. It also offers an interesting point of comparison between SL models and spacetime discreteness.

The advantage of working in the ultra-relativistic limit is that the forward equation (\ref{FWD}) can be solved. With initial condition $p_0(E|E_0) = \delta(E-E_0)$  the probability distribution for $E$ at time $t$ conditional on a value $E_0$ at time $0$ is found to be \cite{CIR}
\begin{align}
	p_t(E|E_0) = \frac{\alpha}{\beta} \frac{E}{E_0} e^{-\alpha(E-\beta E_0)}
				I_2\left( 2\alpha\sqrt{\beta E E_0} \right)
\label{Edist}
\end{align}
where
\begin{align}
	\alpha &= \frac{\dot{a}}{a}\frac{m}{D}\frac{1}{1-\beta},  \\
	\beta &= \exp\left\{ -\frac{\dot{a}}{a} t\right\}, 
\end{align}
and $I_2$ is a modified Bessel function of the first kind of order 2. The expected value of $E$ at time $t$ is
\begin{align}
	\mathbb{E}_t[E|E_0]=\beta E_0+\frac{3}{\alpha},
	\label{een}
\end{align}
and the variance is 
\begin{align}
	{\rm Var}_t[E|E_0] = \frac{2\beta}{\alpha}E_0+\frac{3}{\alpha^2}.
\end{align}

We can also write down the asymptotic limit of (\ref{Edist}) valid when $\alpha\sqrt{\beta E E_0}\rightarrow \infty$ 
\begin{align}
	p_t(E|E_0) \propto \sqrt{\frac{m}{4\pi Dt}}\left(\frac{E^{3}}{E_0^{5}}\right)^{1/4}
		\exp\left\{-\frac{m}{Dt}\left(\sqrt{E}-\sqrt{E_0}\right)^2\right\},
\end{align}
where we have assumed that $t\ll a/\dot{a}$. And for completeness we give the steady state distribution (despite the fact that this state is not achieved in the lifetime of the Universe)
\begin{align}
	p_{\infty}(E) = \frac{1}{2}\omega^3E^2 e^{-\omega E'},
	\label{jut}
\end{align}
where 
\begin{align}
	\omega = \frac{\dot{a}}{a}\frac{m}{D}.
\end{align}
The mean of the steady state distribution is $3/\omega$ and the variance is $3/\omega^2$. Eq.(\ref{jut}) is in fact the ultra-relativistic limit of the Maxwell-J\"uttner distribution describing the energies of particles in an ideal gas in thermal equilibrium at relativistic temperatures.

\begin{figure}[h]
        \begin{center}
        	\includegraphics[width=16cm]{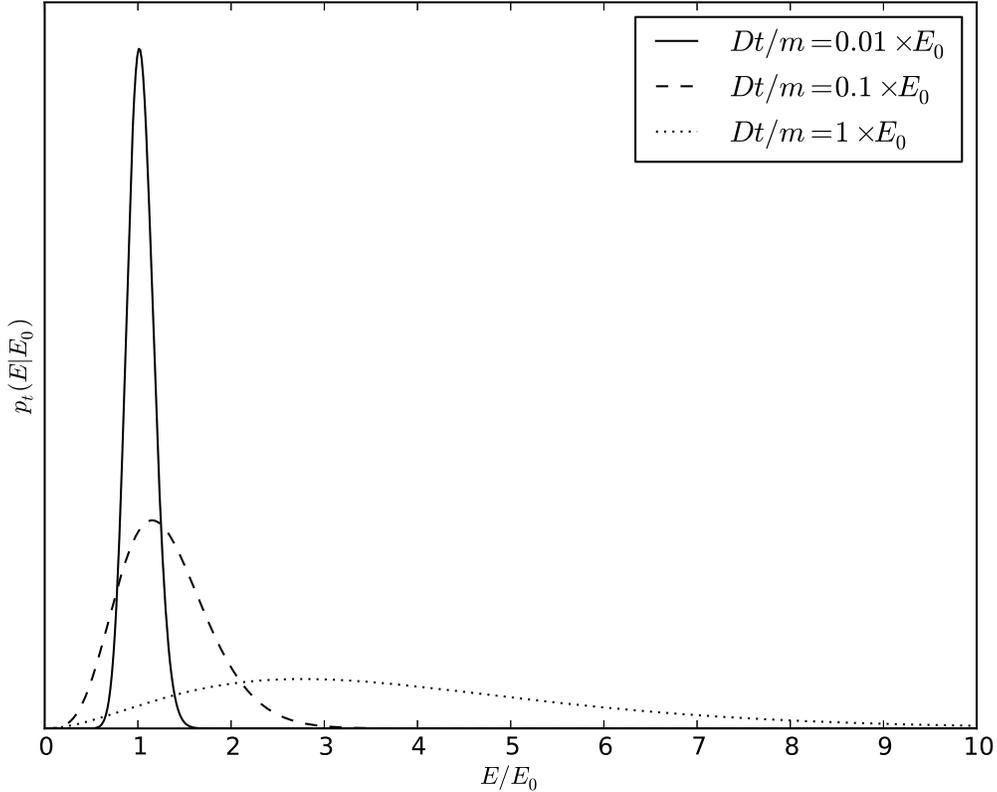}
        \end{center}
\caption{Probability distribution of particle energy. See text for detailed description.}
\label{F1}
\end{figure}

In Fig.\ref{F1} we show the probability distribution for $E$ given in Eq.(\ref{Edist}). Once we fix units such that $E_0=1$ and make the assumption that $t \ll a/\dot{a}$, the shape of the distribution depends only on the value of the combination $Dt/m$. For large values ($>E_0$) the distribution is wide and flat whilst for small values ($<E_0$) the distribution becomes sharp and narrow tending towards a delta function at $E=E_0$ as $Dt/m\rightarrow 0$.

From Eq.(\ref{een}) we see that the expected increase in energy is approximately $3Dt/m$. We can estimate $Dt/m$ using the GRW parameters along with (\ref{D0}) and (\ref{lamb}). If we generously assume that the particle has been freely travelling for almost the whole lifetime of the Universe, $t\sim 10^{17}s$,  
\begin{align}
\left.\frac{Dt}{m}\right|_{GRW} \sim 10^{-15} mc^2.
\end{align}
We can make a more aggressive estimate by choosing $\lambda\alpha\sim 10^{8}m^{-2}s^{-1}$ - this is the order of the current upper bound (CUB) on SL parameters imposed by diffraction experiments using large molecules \cite{TUM2}. Here using the same value for $t$ we find
\begin{align}
\left.\frac{Dt}{m}\right|_{CUB} \sim 10^{-5} mc^2.
\end{align}
In both cases the value is very small. Given our assumption that the particle is travelling with speed close to the speed of light its initial energy $E_0$ must be at least several times the rest energy. Since $E_0\gg Dt/m$ the distribution of energies will be highly peaked around $E_0$ (see Fig.\ref{F1}) with variance of order $DtE_0/m$. Given the fact that it would be difficult to identify a source of massive free particles in the early Universe with very precise energy we conclude that it is very unlikely that this effect could be measured.

\section{Summary and conclusions}
\label{S4}

We have considered the case of a relativistically-fast moving particle with fixed initial energy traversing the Universe over billions of years. Standard quantum theory predicts that the energy of the particle remains fixed and that the wave function slowly disperses in space. By contrast the CSL model, a modification of standard quantum theory to include quantum state reduction as a dynamical process, predicts that the wave function remains localized and that the energy of the particle undergoes diffusion. We have performed a calculation to determine the distribution of possible kinetic energies obtained by the relativistic particle after a long period of CSL evolution.

Starting with the non-relativistic CSL model we have demonstrated that in the case of a sufficiently-localized single particle the stochastic equations for the state vector can be recast in the simplified form of quantum mechanics with universal position localization (QMUPL). The properties of QMUPL are well understood. In particular there is a steady state form for the wave packet achieved after a finite amount of free propagation. The average position and momentum of the packet then satisfy a closed pair of coupled classical diffusion equations. This essentially reduces the complex quantum/stochastic behaviour of CSL to the simple problem of a classical diffusion in phase space.

Although the non-relativistic CSL equations cannot be taken to apply for relativistic systems we argued that they should be valid in the particle's rest frame. This was based on examining the form of proposed relativistic extensions of SL along with the general principle that relativistic SL models should somehow incorporate a fixed localization length scale. The obvious way to do this without the use of a preferred frame or foliation is to make reference to local rest frames invariantly defined by the state. Roughly speaking this means that the localizations of a particle's wave function will happen in the rest frame.

We showed how the classical diffusion process satisfied by the steady state wave packet in the particle rest frame can be Lorantz transformed to describe the diffusion from the point of view of an inertial frame. This led us to derive a forward equation for the observed probability distribution of energies in the case where the inertial observer sees the particle with speed $v\sim c$. The energy process in this case is an example of a CIR process. These are well known from finance where they are useful for describing short rates. The forward equation has a solution which we presented. In particular we were able to show that the solution does not permit negative energies as we would expect.

In fitting estimates for the CSL parameters to this energy distribution we found that even using the upper bound values obtained from diffraction experiments, the spread in energy and the average increase in energy due to the diffusion were both very small when compared to the initial kinetic energy of the particle. Given that the initial energy of the particle in practical examples will have some uncertainty, it is unlikely that the precision required to measure the relativistic energy diffusion due to CSL can be achieved. On the other hand this result means that the energy increases due to CSL are kept small, even on the scale of the lifetime of the Universe and therefore do not pose a problem for the viability of the theory.

Given that the localizations which happen in the rest frame are Lorentz contracted from the perspective of a fast moving observer, one might have expected that the collapse effects would be stronger for relativistic particles. However, this has to be balanced against the time dilation effects which effectively reduce the localization rate from the observer's point of view. On the basis of the above calculation the two effects seem to cancel each other out. Relativistic particles do not obviously provide a way to amplify the effects of CSL for the purpose of experimental test.

\section*{Acknowledgements}
I would like to thank Carlo Contaldi, Arttu Rajantie, Fay Dowker, and Philip Pearle for helpful discussions and comments.

\end{document}